\documentclass[prd,aps,showpacs,nofootinbib,tightenlines]{revtex4-1}

\usepackage{epsfig}
\usepackage[latin1]{inputenc}
\usepackage{float,amsmath,color}
\usepackage{graphicx}

\newcommand{\beq}{\begin{equation}}
\newcommand{\eeq}{\end{equation}}
\newcommand{\bqa}{\begin{eqnarray}}
\newcommand{\eqa}{\end{eqnarray}}

\def\lsim{\mathrel{\rlap{\lower4pt\hbox{$\sim$}}
    \raise1pt\hbox{$<$}}}                
\def\gsim{\mathrel{\rlap{\lower4pt\hbox{$\sim$}}
    \raise1pt\hbox{$>$}}}                

\begin{document}

\title{A brief overview of hard-thermal-loop perturbation theory}

\author{Nan Su}
\email{nansu@physik.uni-bielefeld.de}
\affiliation{Fakult\"at f\"ur Physik, Universit\"at Bielefeld, D-33615 Bielefeld, Germany}

\begin{abstract}

The poor convergence of quantum field theory at finite temperature has been one of the main obstacles in the practical applications of thermal QCD for decades. Here we briefly review the progress of hard-thermal-loop perturbation theory (HTLpt) in reorganizing the perturbative expansion in order to improve the convergence. The quantum mechanical anharmonic oscillator is used as a simple example to show the breakdown of weak-coupling expansion, and variational perturbation theory is introduced as an effective resummation scheme for divergent weak-coupling expansions. We discuss HTLpt thermodynamic calculations for QED, pure-glue QCD, and QCD with $N_f=3$ up to three-loop order. The results suggest that HTLpt provides a systematic framework that can be used to calculate both static and dynamic quantities for temperatures relevant at LHC.

\end{abstract}

\maketitle

\tableofcontents

\section{Introduction}
\label{intro}

Since the beginning of experiments in 1999, the Relativistic Heavy-Ion Collider (RHIC) at Brookhaven National Laboratory has been doing intensive studies at initial temperatures up to twice the critical temperature for deconfinement,\footnote{Note that the deconfinement transition is actually an analytic crossover \cite{Aoki:2006we} and $T_c$ represents a temperature around which the thermodynamic quantities change quickly.} $T_c \sim170\,$MeV. This translates to a strong coupling constant of $\alpha_s(\mu=2\pi\times170\,{\rm MeV}) \equiv g^2/(4\pi) \sim 0.4$, or equivalently $g\sim2$, where $\mu$ is the renormalization scale and it relates to the temperature by $\mu=2\pi T$. The upcoming experiments at the Large Hadron Collider (LHC) at European Organization for Nuclear Research are expected to yield initial temperatures of $4-6\,T_c$, driving the running coupling further down. For the RHIC and LHC experiments to have the greatest possible impact on science, it is essential to make as close a connection to the fundamental theory of quantum chromodynamics (QCD) as possible. There is an urgent need for theoretical analysis that is based rigorously on QCD but which can also make contact with more phenomenological approaches, particularly in the area of thermodynamics and real-time dynamics of QCD at \emph{intermediate coupling}, $g\sim2$. 

We have to be extremely careful when dealing with this intermediately coupled region. Naively, $g\sim2$ seems to suggest the breakdown of perturbation theory in this region. This is also in line with the observations from the early RHIC data that the state of matter created there behaved more like a strongly coupled fluid than a weakly coupled plasma~\cite{Arsene:2004fa,Back:2004je,Adams:2005dq,Adcox:2004mh,Gyulassy:2004zy}. As a result, the term ``quark-gluon plasma'' might need to be modified to ``quark-gluon liquid'', and a description in terms of strong coupling formalisms, such as hydrodynamics or AdS/CFT correspondence, might be more appropriate. However on the other hand, $g\sim2$ is not huge especially when considering that $\alpha_s=g^2/(4\pi)\sim0.4$ is still a small number. So people have not yet totally lost faith on perturbation theory and as a payback observables like jet quenching~\cite{Qin:2007rn,Qin:2009gw} and elliptic flow~\cite{Xu:2007jv} have been able to be described using a perturbative formalism. Therefore it seems that a complete understanding of QGP would require knowledge from both strong-coupling and weak-coupling formalisms, and in this paper we would focus on the latter approach.

Thermodynamics describes the bulk properties of matter in or near equilibrium which are theoretically clean and well defined. The calculation of thermodynamic functions for finite temperature field theories has a long history. In the early 1990s the free energy was calculated to order $g^4$ for massless scalar $\phi^4$ theory \cite{Frenkel:1992az,Arnold:1994ps}, quantum electrodynamics (QED) \cite{Arnold:1994eb} and QCD \cite{Arnold:1994ps,Arnold:1994eb}, respectively. The corresponding calculations to order $g^5$ were obtained soon afterwards \cite{Parwani:1994zz, Braaten:1995cm,Parwani:1994je,Parwani:1994xi,Andersen:1995ej,Zhai:1995ac,Braaten:1995ju,Braaten:1995jr}. Recent results have extended the calculation of the QCD free energy by determining the coefficient of the $g\log g$ contribution \cite{Kajantie:2002wa}. For massless scalar theories the perturbative free energy is now known to order $g^6$ \cite{Gynther:2007bw} and $g^8 \log g$ \cite{Andersen:2009ct}.

\begin{figure}[t]
\begin{center}
\includegraphics[width=0.5\textwidth]{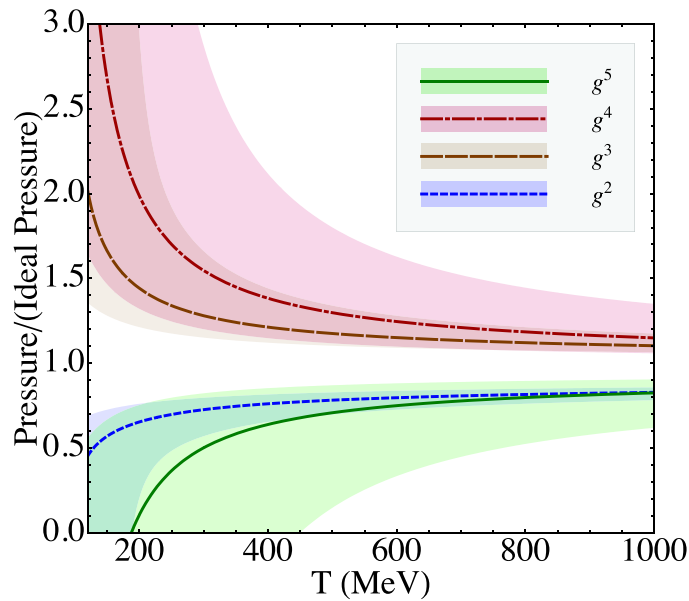}
\end{center}
\caption{Weak-coupling expansion for the scaled QCD pressure with $N_f=3$. Shaded bands show the result of varying the renormalization scale $\mu$ by a factor of 2 around $\mu=2\pi T$.}
\label{fig:pertpressure-qcd}
\end{figure}

\begin{figure}[h]
\begin{center}
\includegraphics[width=0.5\textwidth]{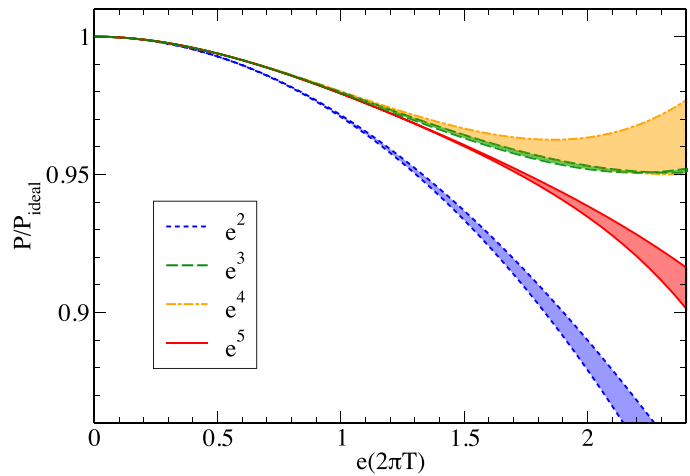}
\end{center}
\caption{Weak-coupling expansion for the scaled QED pressure with $N_f=1$. Shaded bands show the result of varying the renormalization scale $\mu$ by a factor of 2 around $\mu=2\pi T$.}
\label{fig:pertpressure-qed}
\end{figure}

Unfortunately, for all the above mentioned theories the resulting weak-coupling approximations, truncated order-by-order in the coupling constant, are poorly convergent and show large dependence on the renormalization scale unless the coupling constant is tiny which corresponds to astronomically high temperatures. In Fig.~\ref{fig:pertpressure-qcd}, we show the weak-coupling expansion for the QCD pressure with $N_f=3$ normalized to that of an ideal gas through order $g^5$. The various approximations oscillate wildly and show no signs of convergence in the temperature range shown which is probed in the ongoing experiments. The bands are obtained by varying the renormalization scale $\mu$ by a factor of 2 around the value $\mu=2\pi T$ and we use three-loop running for $\alpha_s$~\cite{Amsler:2008zzb} with $\Lambda_{\overline{\rm MS}}(N_f=3)=344\,$MeV~\cite{McNeile:2010ji}. In Fig.~\ref{fig:pertpressure-qed} we show the weak-coupling expansion for the QED pressure with $N_f=1$ normalized to that of an ideal gas through order $e^5$, and we see clearly the same poor convergence pattern as the QCD case. Therefore this oscillating behavior is not specific to QCD, but a generic feature for hot field theories which actually has been also observed in scalar theories~\cite{Frenkel:1992az,Arnold:1994ps,Parwani:1994zz, Braaten:1995cm,Gynther:2007bw,Andersen:2009ct}. Due to this subtlety, a straightforward perturbative expansion in powers of $\alpha_s$ for QCD does not seem to be of any quantitative use even at temperatures many orders of magnitude higher than those achievable in heavy-ion collisions. 

The poor convergence of finite-temperature perturbative expansions of thermodynamic functions stems from the fact that at high temperature the classical solution is not described by massless particle states. Instead one must include plasma effects such as the screening of electric fields and Landau damping via a self-consistent hard-thermal-loop (HTL) resummation~\cite{Braaten:1989mz}. The inclusion of plasma effects can be achieved by reorganizing perturbation theory. There are several ways of systematically reorganizing the finite-temperature perturbative expansion~\cite{Blaizot:2003tw,Kraemmer:2003gd,Andersen:2004fp}. In this paper we will focus on the hard-thermal-loop perturbation theory (HTLpt) method~\cite{Andersen:1999fw,Andersen:1999sf,Andersen:1999va,Andersen:2002ey,Andersen:2003zk,Andersen:2009tw,Andersen:2009tc,Andersen:2010ct,Andersen:2010wu,Andersen:2011sf,Andersen:2011ug,Su:2011zv}. The HTLpt method is inspired by variational perturbation theory (VPT)~\cite{Kleinert93,JK93,Kleinert:1993am,Karrlein:1993ej,Janke:1995wu,Janke:1995zz,Kleinert:1995hc,Kleinert-book-95}. HTLpt is a gauge-invariant extension of screened perturbation theory (SPT)~\cite{Karsch:1997gj,Chiku:1998kd,Andersen:2000yj,Andersen:2001ez,Andersen:2008bz}, which is a perturbative reorganization for finite-temperature massless scalar field theory. In the SPT approach, one introduces a single variational parameter which has a simple interpretation as a thermal mass. In SPT a mass term is added to and subtracted from the scalar Lagrangian, with the added piece kept as part of the free Lagrangian and the subtracted piece associated with the interactions. The mass parameter is then required to satisfy either a variational or perturbative prescription. This naturally leads to the idea that one could apply a similar technique to gauge theories by adding and subtracting a mass in the Lagrangian. However, in gauge theories, one cannot simply add and subtract a local mass term since this would violate gauge invariance. Instead, one adds and subtracts an HTL effective action which modifies the propagators and vertices selfconsistently so that the reorganization is manifestly gauge invariant~\cite{Braaten:1991gm}.

This paper focuses on the resummation of thermodynamics for gauge theories. In Sec.~\ref{anhar} we use a one-dimensional quantum mechanical example, namely the anharmonic oscillator, to demonstrate the breakdown of weak-coupling expansion in a simple setup. Then in Sec.~\ref{VPT} we introduce variational perturbation theory which is an effective resummation scheme for divergent weak-coupling expansions. Hard-thermal-loop perturbation theory is introduced in Sec.~\ref{HTLpt}, where we discuss its formalism and results of thermodynamic calculations to three-loop order for QED, pure-glue QCD, and QCD with $N_f=3$. We conclude in Sec.~\ref{concl} with a brief outlook for the real-time and other applications of HTLpt.

\section{Anharmonic oscillator}
\label{anhar}

Before entering the complicated field theory, let us make our life simple by first going back to quantum mechanics. In this section we will use an example of one-dimensional anharmonic oscillator to show that weak-coupling expansion totally breaks down even for this simple system no matter how small the coupling is, in contrast to daily intuition.

\begin{figure}[t]
\begin{center}
\includegraphics[width=0.5\textwidth]{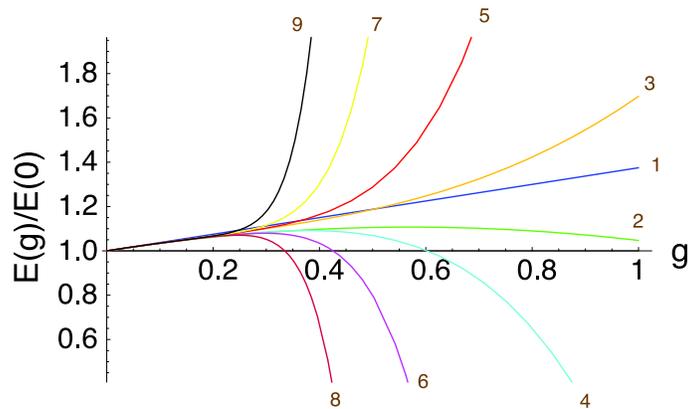}
\end{center}
\caption{Weak-coupling expansion results of the ground-state energy of an anharmonic oscillator.}
\label{fig:anharmonic1}
\end{figure}

Now let us consider the ground-state energy $E$ of a simple anharmonic oscillator with potential
\beq
V(x) = {1\over2} \omega^2 x^2 + {g\over4}x^4\;,
\label{V_anharmonic}
\eeq
with $\omega$ and $g$ some arbitrary positive coefficients. The weak-coupling expansion of the ground-state energy has been calculated to all orders by Bender and Wu~\cite{Bender:1968sa,Bender:1969si,Bender:1971gu,Bender:1990pd} and the result reads
\beq
E(g) = \omega \sum_{n=0}^\infty c_n^{\rm BW} \left({g\over4\omega^3}\right)^n\;,
\label{E_BW}
\eeq
where $c_n^{\rm BW}$ are rational coefficients obtainable to all orders
\beq
c_n^{\rm BW} = \left\{{1\over2},{3\over4},-{21\over8},{333\over16},-{30885\over16},\ldots\right\}\;.
\eeq
In the limit $n\rightarrow\infty$, $c_n^{\rm BW}$ becomes
\beq
\lim_{n\rightarrow\infty}c_n^{\rm BW} = (-1)^{n+1} \sqrt{6\over\pi^3} 3^n (n-{1\over2})!\;.
\label{c_inf}
\eeq
Due to the factorial growth in Eq.~(\ref{c_inf}), the weak-coupling expansion ground-state energy in Eq.~(\ref{E_BW}) is an asymptotic series with zero radius of convergence. The resulting weak-coupling expansion approximations to the ground-state energy are plotted in Fig.~\ref{fig:anharmonic1}. The vertical axis is the anharmonic oscillator ground-state energy scaled by the corresponding harmonic oscillator ground-state energy $E(g)/E(0)$, and the horizontal axis is the coupling strength $g$. Curves in different colors correspond to the ground-state energy up to various orders in the weak-coupling expansion. Instead of converging, Fig.~\ref{fig:anharmonic1} shows clearly that the weak-coupling results are oscillating. If keep adding higher order terms, the results would be bended up and down more steeply. Finally as $n\rightarrow\infty$, the resulting curve would blow to infinity right from the origin which demonstrates the meaning of zero radius of convergence. This result is striking and highly nontrivial: Intuitively ``a small coupling expansion'' sounds like ``a perturbative expansion'', however the above result actually indicates that ``small coupling'' may not be ``perturbative'' which is totally against our intuition! The anharmonic oscillator therefore provides a simple example for the breakdown of weak-coupling expansion, and calls the need of resummation.

\section{Variational perturbation theory}
\label{VPT}

In order to improve the convergence of the weak-coupling expansion and with the inspiration from the Feynman-Kleinert variational approach to path integral~\cite{Feynman:1986ey},\footnote{See also the independent development by Giachetti and Tognetti in Refs.~\cite{GT85-PRL,GT86,GT86-PRB}} 
strong-coupling expansion of variational perturbation theory was developed in the 1990s~\cite{Kleinert93,JK93,Kleinert:1993am,Karrlein:1993ej,Janke:1995wu,Janke:1995zz,Kleinert:1995hc,Kleinert-book-95}.\footnote{For the development and other variants of VPT, we refer the reader to Refs.~\cite{Yukalov76-1,Yukalov76-2,Caswell:1979qh,Halliday:1979vn,Seznec:1979ev,Barnes:1978cd,Killingbeck81,Stevenson:1981vj,Stevenson:1982wn,LGZJ83,Shaverdian:1983ay,Stevenson:1982qw,Yamazaki:1983he,Mitter:1983nt,Stevenson:1984rt,Stevenson:1985zy,Stevenson:1986bq,Okopinska:1987hp,Okopinska:1986pd,Duncan:1988hw,Namgung:1989xh,Polley:1989wf,Thoma:1989in,Ritschel:1989ib,Jones:1989df,Stancu:1989sk,Neveu:1989ny,Ritschel:1990uh,Yukalov91,Gandhi:1990yj,Tarrach:1991ev,Haugerud:1991nh,Bender:1991wa,Gandhi:1992xd,Buckley:1992pc,Duncan:1992ba,Yamada:1992xg,Klimenko:1992av,Sisakian:1993uy,Sisakian:1994xr,Bender:1993nd,Arvanitis:1995em,Guida:1994zv,Bellet:1994mf,Bellet:1994jz,Buchmuller:1994qy,Alexanian:1995rp,Arvanitis:1995bs,Arvanitis:1996pd,Jackiw:1997jga,Cornwall:1997dc} for a far from complete list of early references.}

The basic idea of VPT is rather simple: First, the harmonic term of the potential is split into a new harmonic term with a trial frequency $\Omega$ and a reminder:
\beq
\omega^2 \rightarrow \Omega^2 + \left(\omega^2-\Omega^2\right)\;.
\eeq
Then the anharmonic potential is rewritten into
\beq
V(x) = {\Omega^2 \over 2}x^2+V_{\rm int}(x),
\eeq
with an interaction
\beq
V_{\rm int}(x) = {g \over 4}(rx^2+x^4)\;,
\eeq
where
\beq
r \equiv {2\over g}\left(\omega^2-\Omega^2\right)\;.
\eeq
After this rewriting, a perturbation expansion is carried out at fixed $r$ which generates the new ground-state energy:
\beq
E_N(g,r) = \Omega \sum_{n=0}^N c_n(r) \left({g\over4\Omega^3}\right)^n\;,
\label{E_VPT}
\eeq
where the new coefficients $c_n(r)$ are obtained from the old Bender-Wu coefficients $c_n^{\rm BW}$ through
\beq
c_n(r) = \sum_{j=0}^n c_j^{\rm BW} \left(
\begin{array}{c}
(1-3j)/2 \\
n-j\\
\end{array}
\right)
	\left( 2 r \Omega \right)^{n-j}\;.
\eeq
Recall here that the trial frequency $\Omega$ was introduced solely as an auxiliary parameter for computational convenience. It is not in the original anharmonic potential~(\ref{V_anharmonic}), therefore it should not appear in the final result. The $\Omega$ dependence of Eq.~(\ref{E_VPT}) can be eliminated by requiring the principle of minimal sensitivity at each order $N$
\beq
{\partial E_N \over \partial \Omega}\bigg|_{\Omega=\Omega_N} = 0\;.
\eeq

\begin{figure}[t]
\begin{center}
\includegraphics[width=0.5\textwidth]{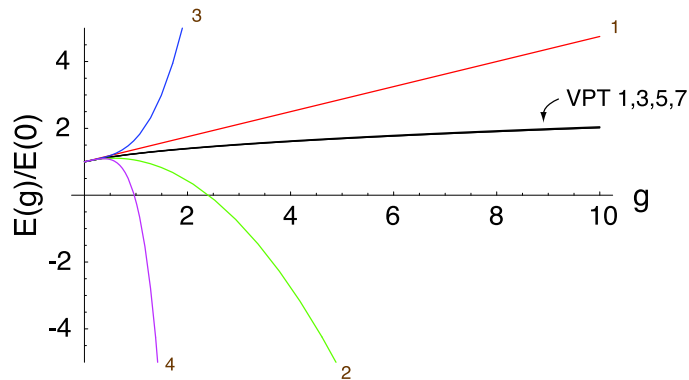}
\end{center}
\caption{Variational perturbation theory results of the ground-state energy of an anharmonic oscillator.}
\label{fig:anharmonic2}
\end{figure}

The VPT ground-state energy is plotted in Fig.~\ref{fig:anharmonic2}. The axes are the same as in Fig.~\ref{fig:anharmonic1}. The black curves are the VPT results up to the first four odd orders in the new expansion. They are almost completely overlapped with each other, and one cannot distinguish the difference between different orders by eyes.\footnote{The figure is generated by Mathematica for illustration. Due to numerical subtleties, the results up to the first two even orders are not easily obtained in Mathematica, therefore the even orders are skipped in Fig.~\ref{fig:anharmonic2}.} However despite of the apparent convergence, we still have to ask whether the VPT results converge to the correct value. By performing a strong-coupling expansion, It has been shown by Janke and Kleinert~\cite{Janke:1995zz} that the VPT results for the the first strong-coupling expansion coefficient $\alpha_0$ agreed to all 23 digits with the most accurate value for $\alpha_0$ available in the literature~\cite{VC91}. The convergence radius of the VPT strong-coupling expansion was rigorously proven to be infinity in Ref.~\cite{Guida:1995px}. 

We have been convinced by VPT in improving the convergence of the weak-coupling expansion from the above simple example. The natural question here to ask is whether such trick could be applied to field theory, especially to QCD for phenomenological interest. The answer is yes, but we have to do it very carefully, cause simply adding a mass term to the QCD Lagrangian would violate gauge symmetry even at the Lagrangian level. We are going to address this subtlety in the next section.

\section{Hard-thermal-loop perturbation theory}
\label{HTLpt}

We have seen the oscillating behavior of weak-coupling expansion for thermal gauge theories in Sec.~\ref{intro}. In order to improve the convergence of perturbation calculations and with the inspiration from VPT, hard-thermal-loop perturbation theory was introduced by Andersen, Braaten and Strickland as a reorganization of thermal gauge theory in 1999 and the one-loop or leading order (LO) thermodynamic calculations were carried out in Refs.~\cite{Andersen:1999fw,Andersen:1999sf,Andersen:1999va}. Later on HTLpt got extended to two loops or next-to-leading order (NLO) in which a further mass expansion was introduced to make the calculation tractable in practice~\cite{Andersen:2002ey,Andersen:2003zk}. The three-loop or next-to-next-to-leading order (NNLO) HTLpt calculations have been accomplished recently for both Abelian~\cite{Andersen:2009tw} and non-Abelian~\cite{Andersen:2009tc,Andersen:2010ct,Andersen:2010wu,Andersen:2011sf} theories, and the NNLO results turned out to be completely analytic. In the rest of this section, we are going to discuss the setup of HTLpt, and results of thermodynamic calculations through NNLO. Please check the above mentioned references for calculational details.

\subsection{Formalism}

In the following of this subsection we will use QCD as an example to show the setup of HTLpt, however the result is general for both Abelian and non-Abelian gauge theories as we will comment in the next subsections.

The Lagrangian density for QCD in Minkowski space reads
\bqa
{\cal L}_{\rm QCD} =
-{1\over2}{\rm Tr}\left[G_{\mu\nu}G^{\mu\nu}\right]
+i \bar\psi \gamma^\mu D_\mu \psi 
+{\cal L}_{\rm gf}
+{\cal L}_{\rm gh}
+\Delta{\cal L}_{\rm QCD}\;,
\label{L-QCD}
\eqa
where the gluon field strength is $G^{\mu\nu}=\partial^{\mu}A^{\nu}-\partial^{\nu}A^{\mu}-ig[{ A^{\mu},A^{\nu}}]$, the term with the quark fields $\psi$ contains an implicit sum over the $N_f$ quark flavors, and the covariant derivative is $D^{\mu}=\partial^{\mu}-igA^{\mu}$. The ghost term ${\cal L}_{\rm gh}$ depends on the gauge-fixing term ${\cal L}_{\rm gf}$. The perturbative expansion in powers of $g$ generates ultraviolet divergences. The renormalizability of perturbative QCD guarantees that all divergences in physical quantities can be removed by renormalization of the coupling constant $\alpha_s= g^2/(4 \pi)$ and the necessary counterterms are represented by $\Delta{\cal L}_{\rm QCD}$ in the Lagrangian (\ref{L-QCD}). There is no need for wavefunction renormalization, because physical quantities are independent of the normalization of the field. There is also no need for renormalization of the gauge parameter, because physical quantities are independent of the gauge parameter.

HTLpt is a reorganization of the perturbation series for thermal gauge theories with the Lagrangian density written as
\bqa
{\cal L}= \left({\cal L}_{\rm QCD}
+ {\cal L}_{\rm HTL} \right) \Big|_{g \to \sqrt{\delta} g}
+ \Delta{\cal L}_{\rm HTL}.
\label{L-HTLQCD}
\eqa
The HTL improvement term is
\bqa
{\cal L}_{\rm HTL}=-{1\over2}(1-\delta)m_D^2 {\rm Tr}
\left(G_{\mu\alpha}\left\langle {y^{\alpha}y^{\beta}\over(y\cdot D)^2}
	\right\rangle_{\!\!y}G^{\mu}_{\;\;\beta}\right)
         +(1-\delta)\,i m_q^2 \bar{\psi}\gamma^\mu 
\left\langle {y_{\mu}\over y\cdot D}
	\right\rangle_{\!\!y}\psi
	\, ,
\label{L-HTL}
\eqa
where $y^{\mu}=(1,\hat{{\bf y}})$ is a light-like four-vector, and $\langle\ldots\rangle_{ y}$ represents the average over the directions of $\hat{{\bf y}}$. The term~(\ref{L-HTL}) has the form of the effective Lagrangian that would be induced by a rotationally invariant ensemble of charged sources with infinitely high momentum and modifies the propagators and vertices self-consistently so that the reorganization is manifestly gauge invariant~\cite{Frenkel:1989br,Taylor:1990ia,Braaten:1991gm,Efraty:1992pd,Jackiw:1993zr,Jackiw:1993pc}. The parameter $m_D$ can be identified with the Debye screening mass, and $m_q$ with the thermal quark mass to account for the screening effects. HTLpt is defined by treating $\delta$ as a formal expansion parameter. By coupling the HTL improvement term~(\ref{L-HTL}) to the QCD Lagrangian~(\ref{L-QCD}), HTLpt systematically shifts the perturbative expansion from being around an ideal gas of massless particles which is the physical picture of the weak-coupling expansion, to being around a gas of massive quasiparticles which are the more appropriate physical degrees of freedom at high temperature. 

Physical observables are calculated in HTLpt by expanding them in powers of $\delta$, truncating at some specified order, and then setting $\delta=1$. This defines a reorganization of the perturbation series in which the effects of $m_D^2$ and $m_q^2$ terms in~(\ref{L-HTL}) are included to all orders but then systematically subtracted out at higher orders in perturbation theory by the $\delta m_D^2$ and $\delta m_q^2$ terms in~(\ref{L-HTL}), which is in the spirit of VPT. If we set $\delta=1$, the HTLpt Lagrangian (\ref{L-HTLQCD}) reduces to the QCD Lagrangian (\ref{L-QCD}). If the expansion in $\delta$ could be calculated to all orders the final result would not depend on $m_D$ and $m_q$ when we set $\delta=1$. However, any truncation of the expansion in $\delta$ produces results that depend on $m_D$ and $m_q$. Some prescription is required to determine $m_D$ and $m_q$ as a function of $T$ and $\alpha_s$. We will discuss several prescriptions in the next subsections.

The HTL perturbation expansion generates ultraviolet divergences. In QCD perturbation theory, renormalizability constrains the ultraviolet divergences to have a form that can be cancelled by the counterterm Lagrangian $\Delta{\cal L}_{\rm QCD}$. Although the renormalizability of the HTL perturbation expansion has not been proven, the renormalization can be archived through NNLO by using only a vacuum counterterm, a Debye mass counterterm and a fermion mass counterterm, as well as a coupling constant counterterm. The necessary counterterms for the renormalization through NNLO as just discussed read
\bqa
\Delta{\cal E}_0&=&\left({d_A\over128\pi^2\epsilon}
+{\cal O}(\delta\alpha_s)
\right)(1-\delta)^2m_D^4\;,
\label{del1e0}
\\
\Delta m_D^2&=&\left(-{11c_A-4s_F\over12\pi\epsilon}\alpha_s\delta
+{\cal O}(\delta^2\alpha_s^2)
\right)(1-\delta)m_D^2\;,
\label{delmd} \\
\Delta m_q^2&=&\left(-{3\over8\pi\epsilon}{d_A\over c_A}\alpha_s\delta
+{\cal O}(\delta^2\alpha_s^2)
\right)
(1-\delta)m_q^2\;,
\label{delmf}\\
\delta\Delta\alpha_s&=&-{11c_A-4s_F\over12\pi\epsilon}\alpha_s^2\delta^2
+{\cal O}(\delta^3\alpha^3_s)\;,
\label{delalpha}
\eqa
where the coupling constant counterterm is the standard one-loop running of QCD~\cite{Gross:1973id,Politzer:1973fx}.

\subsection{Thermodynamic potentials through NNLO}

The calculation of the free energy in HTLpt involves the evaluation of vacuum diagrams. In Refs.~\cite{Andersen:1999fw,Andersen:1999sf,Andersen:1999va,Andersen:2002ey,Andersen:2003zk,Andersen:2009tw,Andersen:2009tc,Andersen:2010ct,Andersen:2010wu,Andersen:2011sf}, the free energy was reduced to scalar sum-integrals. The one-loop free energy were evaluated exactly by replacing the sums by contour integrals, extracting the poles in $\epsilon$, and then reducing the momentum integrals to integrals that were at most two-dimensional and could therefore be easily evaluated numerically. Evaluating two-loop free energy exactly would involve the evaluation of five-dimensional numerical integrals which turned out to be intractable, and therefore attacking the third loop in this way is hopeless. The fact that $m_{D/q} \sim gT$ suggests that $m_{D/q}/T$ can be treated as expansion parameters of order $g$ in terms of which the sum-integrals can be further expanded~\cite{Andersen:2001ez}. It was shown that the first few terms in the $m_{D/q}/T$ expansion of the sum-integrals gave a surprisingly accurate approximation to the exact result~\cite{Andersen:1999fw,Andersen:1999sf,Andersen:1999va,Andersen:2001ez}. This trick is adopted in the HTLpt results that we are going to present next.

The Feynman diagrams through NNLO in HTLpt are gathered in Appendix~\ref{diagr}. Fig.~\ref{fig:LO&NLO} contains the notation key, as well as the LO and NLO HTLpt vacuum diagrams, while Fig.~\ref{fig:NNLO} shows the NNLO HTLpt vacuum diagrams. Note that all the propagators and vertices in Figs.~\ref{fig:LO&NLO} and \ref{fig:NNLO} are HTL resummed, which is in contrast to the weak-coupling expansion case. The strategy of evaluating the diagrams is to first reduce the diagrams to scalar integrals and then expand the resulting scalar integrals in powers of $m_D/T$ and $m_q/T$ as discussed above. We will carry out the $m_D/T$ and $m_q/T$ expansions to high enough order to include all terms through order $g^5$ if $m_D /T$ and $m_q/T$ are taken to be of order $g$. The two-loop approximation will be in the weak-coupling limit accurate to order $g^3$ and the three-loop approximation accurate to order $g^5$. We will now present the resulting thermodynamic potentials through NNLO and the details of the calculations are presented in Refs.~\cite{Andersen:2002ey,Andersen:2003zk,Andersen:2009tw,Andersen:2010ct,Andersen:2011sf}.

The LO thermodynamic potential reads
\bqa
{\Omega_{\rm LO} \over {\cal F}_{\rm ideal}} = 
1 
+ {7 \over 4}{d_F \over d_A} 
- {15 \over 2} \hat{m}_D^2 
- 30 {d_F \over d_A} \hat{m}_q^2 
+ 30 \hat{m}_D^3
+ {45 \over 4}
\left(\log {\hat \mu \over 2} - {7 \over 2} + \gamma_{ E} + {\pi^2 \over 3} \right) \hat{m}_D^4
- 60 {d_F \over d_A}(\pi^2 - 6) \hat{m}_q^4
\;,
\label{Omega-LO}
\eqa
where ${\cal F}_{\rm ideal}$ is the free energy of a gas of $d_A$ massless spin-one bosons and $\hat{m}_D$, $\hat{m}_q$ and $\hat \mu$ are dimensionless variables:
\bqa
{\cal F}_{\rm ideal} &=& d_A\left(-{\pi^2\over45}T^4\right) \;,
\\
\hat{m}_D &=& {m_D \over 2 \pi T} \;,
\\
\hat{m}_q &=& {m_q \over 2 \pi T} \;,
\\
\hat \mu &=& {\mu \over 2 \pi T} \;. 
\eqa

The NLO thermodynamic potential reads
\bqa
{\Omega_{\rm NLO} \over {\cal F}_{\rm ideal}} &=&
1 
+ {7 \over 4}{d_F \over d_A}
- 15 \hat{m}_D^3 
- {45 \over 4} \left(\log\hat{\mu \over 2} - {7 \over 2} + \gamma_{E} + {\pi^2 \over 3}\right) \hat m_D^4
+ 60{d_F \over d_A}(\pi^2 - 6) \hat{m}_q^4
\nonumber \\ && \nonumber
+ {c_A\alpha_s\over3\pi}
\left[-{15 \over 4} + 45 \hat{m}_D
- {165 \over 4}\left(\log{\hat\mu \over 2} - {36 \over 11}\log\hat{m}_D - 2.001\right) \hat{m}_D^2
+ {495 \over 2}\left(\log{\hat\mu \over 2} + {5 \over 22} + \gamma_E\right) \hat{m}_D^3
\right] 
\\ && \nonumber
+ {s_F\alpha_s \over \pi}\left[-{25 \over 8} + 15\hat{m}_D
+ 5\left(\log{\hat{\mu} \over 2} - 2.33452\right) \hat{m}_D^2
- 30\left(\log{\hat\mu \over 2} - {1 \over 2} + \gamma_E + 2\log2\right) \hat{m}_D^3
\right.\\ &&\left.
- 45\left(\log{\hat\mu \over 2} + 2.19581\right)\hat{m}_q^2
+ 180\hat{m}_D\hat{m}_q^2\right]
\;.
\label{Omega-NLO}
\eqa

The NNLO thermodynamic potential reads
\begin{eqnarray}\nonumber
{\Omega_{\rm NNLO} \over {\cal F}_{\rm ideal}} &=&
1 + {7 \over 4}{d_F \over d_A}
- {15 \over 4}\hat{m}_D^3
+ {c_A\alpha_s \over 3\pi}\left[
- {15 \over 4}
+ {45 \over 2}\hat{m}_D
- {135 \over 2}\hat{m}^2_D
- {495 \over 4}\left( \log{\hat\mu \over 2} + {5 \over 22} + \gamma_E \right)\hat{m}_D^3 \right]
\\ && \nonumber
+ {s_F\alpha_s \over \pi}\left[
- {25 \over 8} + {15 \over 2}\hat{m}_D
+ 15\left( \log{\hat\mu \over 2} - {1 \over 2} + \gamma_E + 2\log2 \right)\hat{m}_D^3
- 90\hat{m}^2_{ q}\hat{m}_D
\right]
\\&& \nonumber
+ \left( {c_A\alpha_s \over 3\pi}\right)^2 \left[ 
{45 \over 4}{1 \over \hat{m}_D}
- {165 \over 8}\left( \log{\hat{\mu} \over 2} 
- {72 \over 11}\log{\hat{m}_D}
- {84 \over 55}
- {6 \over 11}\gamma_E
- {74 \over 11}{\zeta^{\prime}(-1) \over \zeta(-1)}
+ {19 \over 11}{\zeta^{\prime}(-3) \over \zeta(-3)}
\right)
\right.\\ &&\left. \nonumber
+ {1485 \over 4}\left(
\log{\hat{\mu} \over 2} - {79 \over 44} + \gamma_E + \log2 - {\pi^2 \over 11}\right)\hat{m}_D
\right]
+ \left({c_A\alpha_s \over 3\pi}\right)\left({s_F\alpha_s \over \pi}\right)
\left[ 
{15 \over 2}{1 \over \hat{m}_D}
\right.\\&&\left.\nonumber
- {235 \over 16}\left( \log{\hat{\mu} \over 2}
- {144 \over 47}\log{\hat{m}_D}
- {24 \over 47}\gamma_E
+ {319 \over 940}
+ {111 \over 235}\log2
- {74 \over 47}{\zeta^{\prime}(-1) \over \zeta(-1)}
+ {1 \over 47}{\zeta^{\prime}(-3) \over \zeta(-3)}
\right)
\right.\\ &&\nonumber\left.
+ {315 \over 4}\left(\log{\hat{\mu} \over 2} - {8 \over 7}\log2 + \gamma_E + {9 \over 14}
\right)\hat{m}_D
+ 90{\hat{m}_{ q}^2 \over \hat{m}_D}
\right]
\\ &&\nonumber
+ \left({s_F\alpha_s \over \pi}\right)^2
\left[{5 \over 4}{1 \over \hat{m}_D}
+ {25 \over 12}\left(
\log{\hat{\mu} \over 2} + {1 \over 20} + {3 \over 5}\gamma_E - {66 \over 25}\log2
+ {4 \over 5}{\zeta^{\prime}(-1) \over \zeta(-1)}
- {2 \over 5}{\zeta^{\prime}(-3) \over \zeta(-3)}\right)
\right.\\ && \left.
- 15\left(\log{\hat{\mu} \over 2}
- {1 \over 2} + \gamma_E + 2\log2
\right)\hat{m}_D
+ 30{\hat{m}_{ q}^2 \over \hat{m}_D}
\right]
+ s_{2F}\left({\alpha_s \over \pi}\right)^2\left[{15 \over 64}(35 - 32\log2) - {45 \over 2}\hat{m}_D\right]
\;.
\label{Omega-NNLO}
\end{eqnarray}
Note that the NNLO thermodynamic potential is completely analytic, and the coupling constant counterterm listed in Eq.~(\ref{delalpha}) coincides with the known one-loop running of the QCD coupling constant
\beq
\mu \frac{d g^2}{d \mu} = -\frac{(11c_A-4s_F)g^4}{24\pi^2} \;.
\label{runningcoupling}
\eeq
In the next subsections we will present thermodynamic quantities as a function of $g$ evaluated at
the renormalization scale $2 \pi T$.

\subsection{QED thermodynamics}

The thermodynamic potentials for QED are obtained by setting the group factors in Eqs.~(\ref{Omega-LO}), (\ref{Omega-NLO}), and (\ref{Omega-NNLO}) to
\beq
d_A = 1\;,			\qquad
d_F = N_f\;, 		\qquad
c_A = 0\;,			\qquad
s_F = N_f\;,		\qquad
s_{2F} = N_f\;.
\;\;\;\textrm{(QED)}
\eeq

\subsubsection{Mass prescriptions}

The mass parameters $m_D$ and $m_f$ in HTLpt are in principle completely arbitrary, however physically they are proportional to $eT$ at LO in the weak-coupling expansion which has to be satisfied for any mass prescriptions. To complete a calculation, it is necessary to specify $m_D$ and $m_f$ as functions of $e$ and $T$. Two possible mass prescriptions are considered in Ref.~\cite{Andersen:2009tw}: 
\begin{itemize}
\item The variational thermal masses obtained from solving the gap equations
\bqa
{\partial \ \ \over \partial m_D}\Omega(T,\alpha,m_D,m_f,\mu,\delta=1) &=& 0 \;, 
\label{gapmd}\\
{\partial \ \ \over \partial m_f}\Omega(T,\alpha,m_D,m_f,\mu,\delta=1) &=& 0 \;.
\label{gapmf}
\eqa

\item The $e^5$ perturbative Debye mass~\cite{Blaizot:1995kg,Andersen:1995ej} and the $e^3$ perturbative fermion mass~\cite{Carrington:2008dw}
\begin{eqnarray}
\label{mass}
m_D^2 &=& {1 \over 3}N_fe^2T^2\left[
1
- \frac{e^2}{24\pi^2}\left(4\gamma + 7 +4\log \frac{\hat\mu}{2}+8\log2\right)
+ {e^3\sqrt{3} \over 4\pi^3}
\right] \;, \\
m_f^2 &=& {1 \over 8}N_fe^2T^2\left[ 1 - {2.854 \over 4\pi}e \right] \;.
\label{fmass}
\end{eqnarray}
\end{itemize}

\subsubsection{Pressure}

\begin{figure}[t]
\begin{center}
\includegraphics[width=0.45\textwidth]{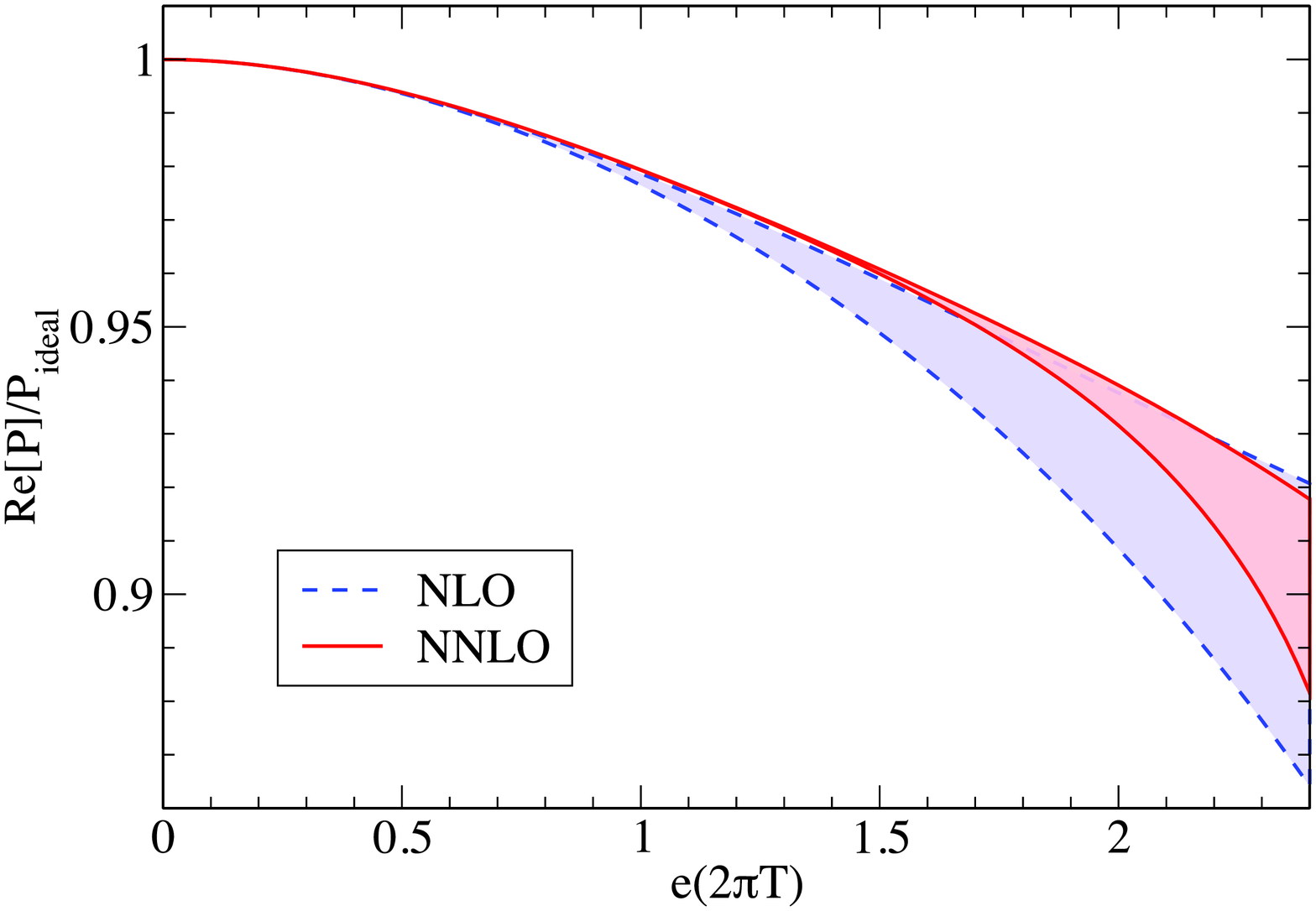}$\;\;\;\;$\includegraphics[width=0.45\textwidth]{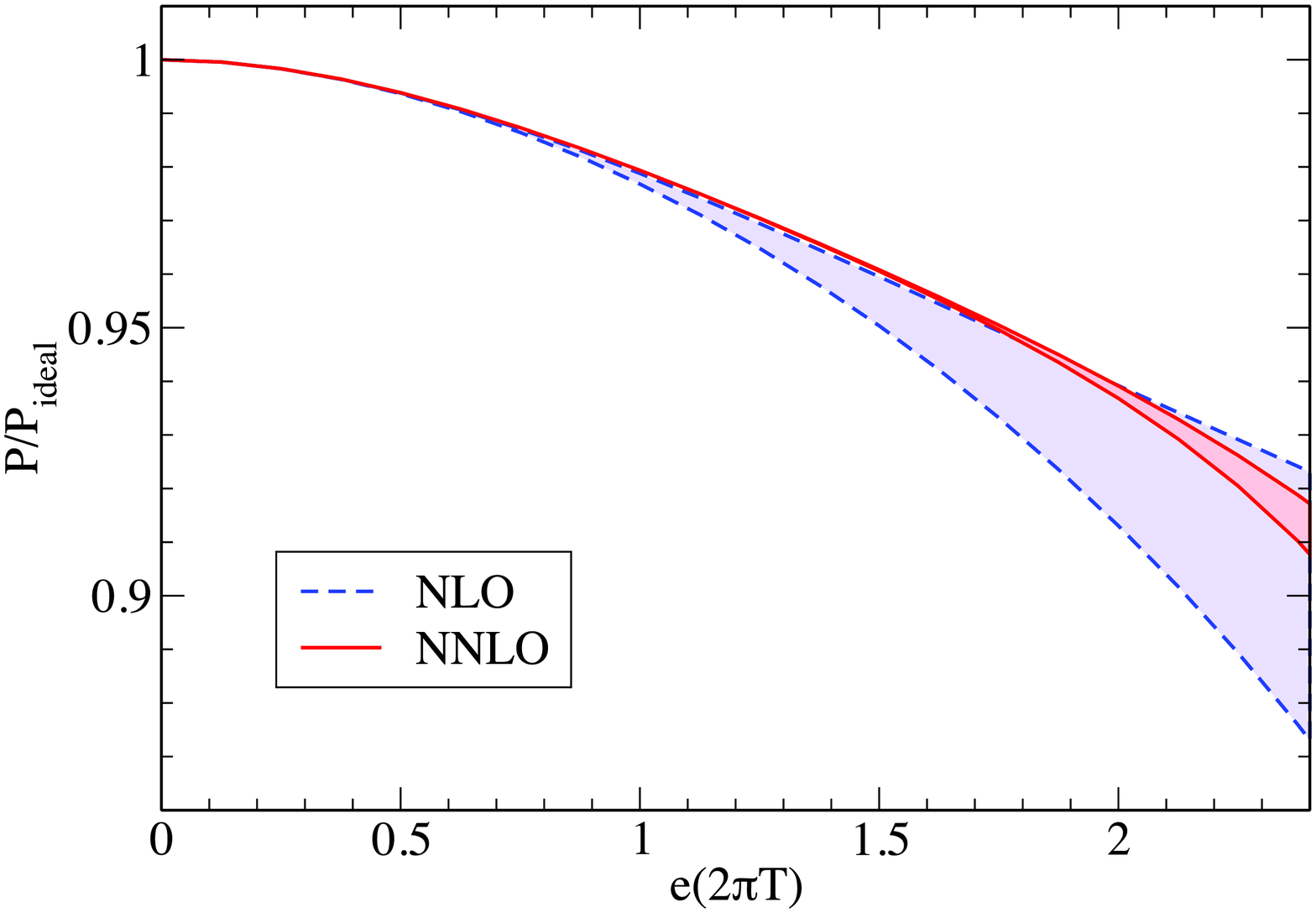}
\end{center}
\caption{\label{fig:NLONNLO} A comparison of the renormalization scale variations between NLO and NNLO HTLpt predictions for the scaled pressure of QED with $N_f=1$ and the variational thermal masses (left panel) and the perturbative thermal masses (right panel). The bands correspond to varying the renormalization scale $\mu$ by a factor of 2 around $\mu=2\pi T$.}
\end{figure}

\begin{figure}[b]
\begin{center}
\includegraphics[width=0.5\textwidth]{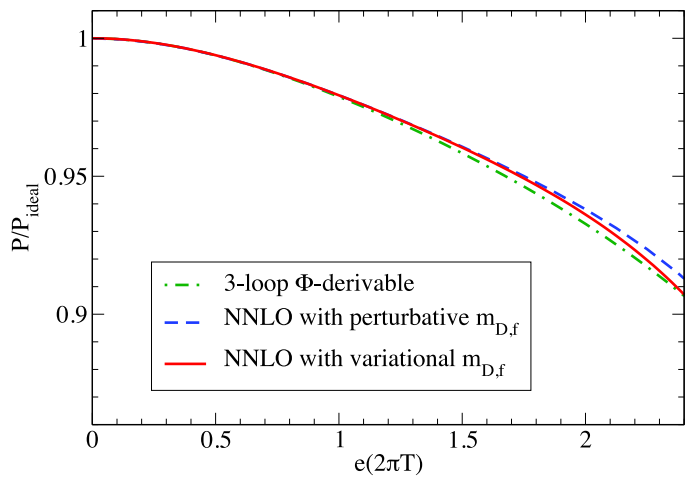}
\end{center}
\caption{A comparison of the predictions for the scaled pressure of QED with 
$N_f=1$ between three-loop $\Phi$-derivable approximation~\cite{Andersen:2004re} and NNLO 
HTLpt at $\mu=2\pi T$.}
\label{fig:PhivsNNLO}
\end{figure}

The resulting predictions for the pressure normalized to that of an ideal gas as a function of $e$ evaluated at the renormalization scale $\mu = 2 \pi T$ are shown in Fig.~\ref{fig:NLONNLO}. Note that when the pressure is evaluated at a scale different than $\mu = 2 \pi T$, we use one-loop running of $e$ to determine the value of the coupling at $\mu = 2 \pi T$. As can be seen from these figures both the variational and perturbative mass prescriptions seem to be consistent when going from NLO to NNLO. At the central value $\mu=2\pi T$, both prescriptions are the same to an accuracy of 0.6\% at $e=2.4$. As a further check, we show a comparison of the NNLO HTLpt results with a three-loop calculation obtained previously using a truncated three-loop $\Phi$-derivable approximation~\cite{Andersen:2004re} in Fig.~\ref{fig:PhivsNNLO}. As can be seen from this figure, the agreement between the NNLO $\Phi$-derivable and HTLpt approaches is at the subpercentage level even at large coupling. The improved convergence for QED gives us confidence to apply the same HTLpt reorganization scheme to QCD.

\subsection{QCD thermodynamics: Pure-glue and $N_f=3$}

The thermodynamic potentials for QCD are obtained by setting the group factors in Eqs.~(\ref{Omega-LO}), (\ref{Omega-NLO}), and (\ref{Omega-NNLO}) to
\beq
d_A = N_c^2-1\;,		\qquad
d_F = N_c N_f\;, 		\qquad
c_A = N_c\;,			\qquad
s_F = {N_f\over2}\;,		\qquad
s_{2F} = {N_c^2-1\over4N_c}N_f\;.
\;\;\;\textrm{(QCD)}
\label{casimirs}
\eeq
The pure-glue QCD results are therefore obtained in the $N_f=0$ limit
\beq
d_A = N_c^2-1\;,		\qquad
d_F = 0\;, 			         \qquad
c_A = N_c\;,			\qquad
s_F = 0\;,		                   \qquad
s_{2F} = 0\;.
 \;\;\;\textrm{(pure-glue)}
\eeq

\subsubsection{Mass prescriptions}

As pointed out earlier, the HTL mass parameters are completely arbitrary and we need a prescription for them in order to complete a calculation. The variational mass prescription unfortunately gives rise to a complex Debye mass and $m_q = 0$ at NNLO. One strategy is therefore to throw away the imaginary part of the thermodynamic potential to obtain thermodynamic functions that are real valued. The problem of a complex mass parameter was encountered at NNLO in SPT massless $\phi^4$ theory~\cite{Andersen:2000yj}, as well as in HTLpt QED~\cite{Andersen:2009tw}, so this seems to be a general issue with SPT/HTLpt at NNLO, which is not specific to QCD. The full variational pressure for QED contains both the real part as shown in the left panel of Fig.~\ref{fig:NLONNLO} and a imaginary part whose absolute value is about 1\% of the real part at large coupling. Since the imaginary part is tiny, therefore the real part provides a good approximation to the pressure as confirmed by the comparison in Fig.~\ref{fig:PhivsNNLO}. However for QCD, the absolute value of the imaginary part of the variational Debye mass becomes comparable to the real part at intermediate coupling, the variational mass prescription is therefore not preferable. Since the weak-coupling perturbative Debye mass receives contributions from the nonperturbative magnetic scale~\cite{Linde:1980ts,Gross:1980br} beyond LO~\cite{Rebhan:1993az,Arnold:1995bh}, for the NNLO HTLpt QCD results that we are going to present, the Braaten and Nieto's (BN) mass parameter of three-dimensional electrostatic QCD (EQCD) will be used as a substitute for the perturbative Debye mass, effectively discarding the nonperturbative contributions. In Ref.~\cite{Braaten:1995ju,Braaten:1995jr} it was calculated to NLO, giving
\bqa
m_D^2&=&
{4\pi\alpha_s\over3}T^2
\left\{
	c_A+s_F
	+{c_A^2\alpha_s\over3\pi}
		\left(
			{5\over 4}+{11\over 2}\gamma_E+{11\over 2}\log{\hat{\mu}\over 2}	
		\right)
	+{c_A s_F \alpha_s\over\pi}
		\left[
			{3\over 4}-{4\over 3}\log 2			
			+{7\over 6}\left(\gamma_E+\log{\hat{\mu}\over 2}\right)
		\right]
	\right.\nonumber\\&&\left.
	+{s_F^2\alpha_s\over\pi}\left({1\over 3}-{4\over 3}\log 2-{2\over 3}\gamma_E-{2\over 3}\log{\hat{\mu}\over 2}\right)
	-{3\over 2}{s_{2F}\alpha_s\over\pi}
\right\}\;.
\label{eq:PmD}
\eqa
The perturbative quark mass does not suffer from the nonperturbative magnetic scale, and it turns out that the final NNLO results are very insensitive to whether one chooses a perturbative mass prescription for $m_q$, or uses the variational mass $m_q=0$. We will therefore use $m_q=0$ for simplicity in the following whenever full QCD is concerned.

\subsubsection{Pressure}

\begin{figure}[t]
\begin{center}
\includegraphics[width=0.46\textwidth]{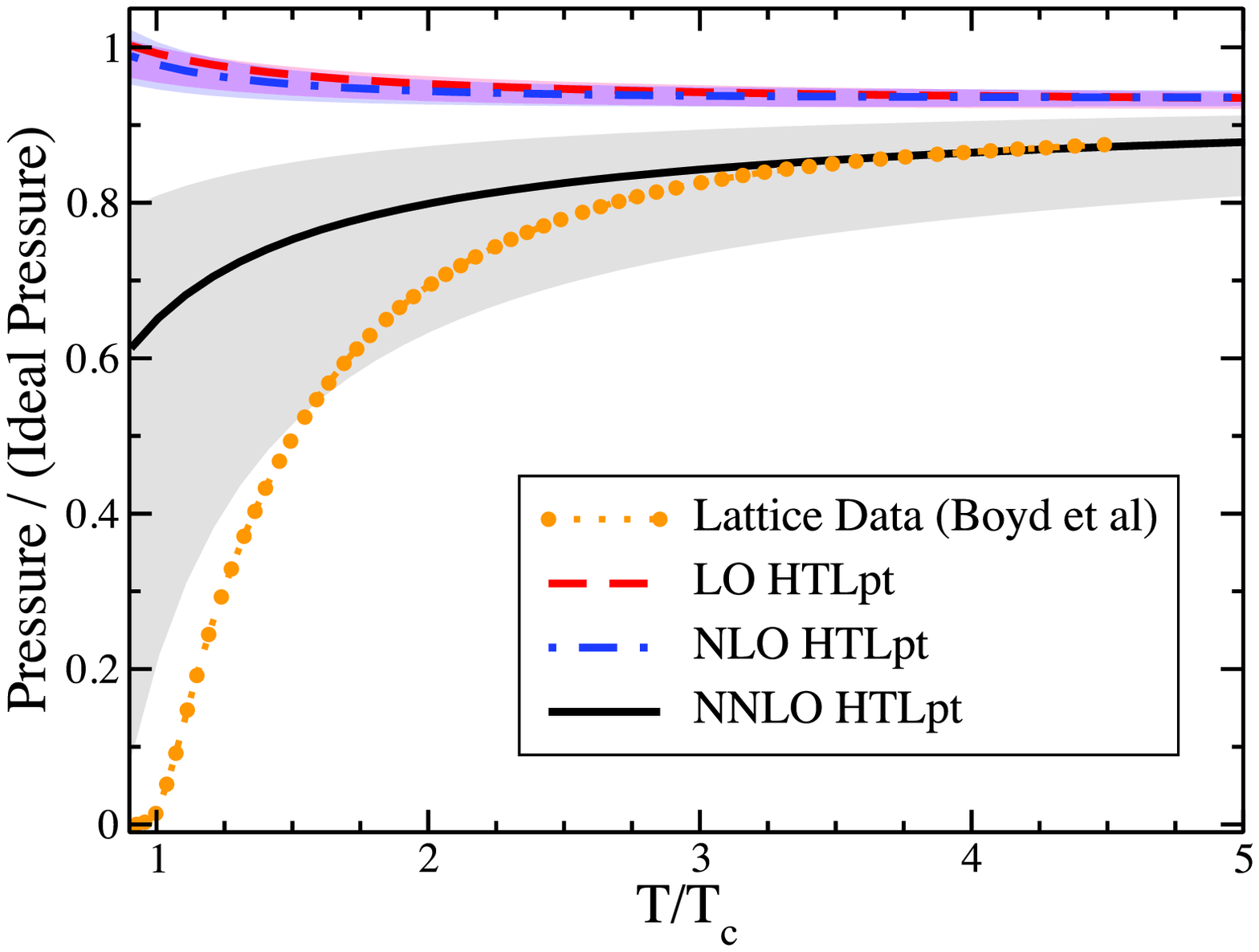}$\;\;\;\;$\includegraphics[width=0.45\textwidth]{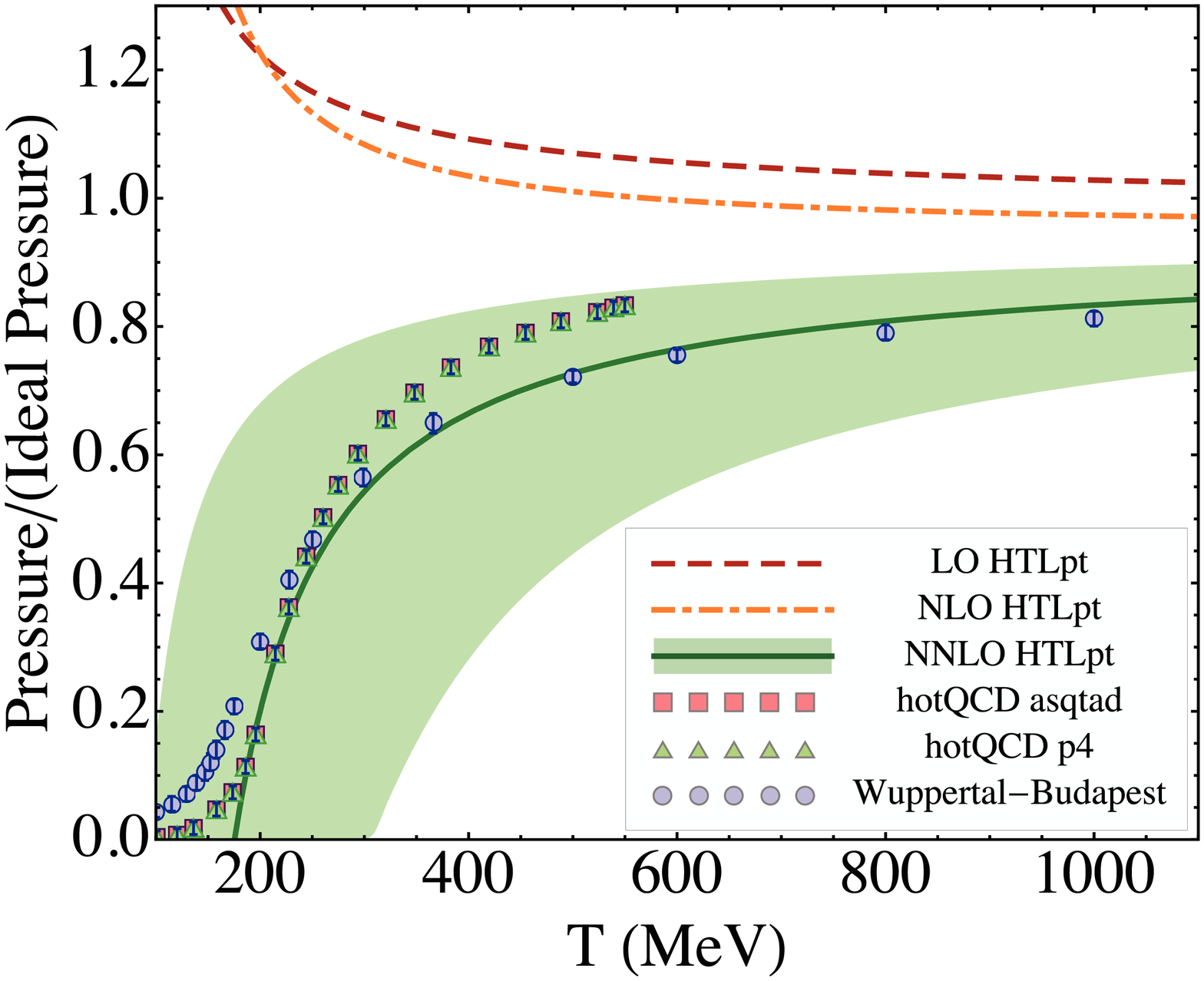}
\end{center}
\caption{\label{fig:pressure} 
Comparison of LO, NLO, and NNLO predictions for the scaled pressure for pure-glue QCD with lattice data from the Bielefeld collaboration~\cite{Boyd:1996bx} (left panel) and $N_f=3$ QCD with $N_f=2+1$ lattice data from the hotQCD~\cite{Bazavov:2009zn} and Wuppertal-Budapest~\cite{Borsanyi:2010cj} collaborations (right panel). Shaded bands show the result of varying the renormalization scale $\mu$ by a factor of 2 around $\mu = 2 \pi T$ for the NNLO result.}
\end{figure}

In the left panel of Fig.~\ref{fig:pressure}, we show the LO, NLO, and NNLO HTLpt predictions for the pressure of pure-glue QCD normalized to that of an ideal gas as a function of $T/T_c$ using the BN mass prescription~(\ref{eq:PmD}) and three-loop running of $\alpha_s$~\cite{Amsler:2008zzb}. The points are lattice data for pure-glue with $N_c=3$ from the Bielefeld collaboration~\cite{Boyd:1996bx}. The bands are obtained by varying the renormalization scale by a factor of 2 around the central value $\mu=2\pi T$. From the plot we see that the convergence of the successive approximations to the pressure is improved over naive perturbation theory. For example, using the naive perturbative approach and comparing the full variation in both successive truncations and renormalization scale variation, one finds that at $T=3\,T_c$ there is variation in the pressure of  $0.69 \leq {\cal P}/{\cal P}_{\rm ideal} \leq 1.32$~\cite{Andersen:2002ey}, whereas using HTLpt there is only a variation of $0.74 \leq {\cal P}/{\cal P}_{\rm ideal} \leq 0.95$. Additionally, at NNLO we see that the $\mu=2\pi T$ result for the pressure in Fig.~\ref{fig:pressure} coincides with the lattice data down to $T \sim 3\,T_c$.

In the right panel of Fig.~\ref{fig:pressure}, we show the LO, NLO, and NNLO HTLpt predictions for the pressure of QCD with $N_f=3$ normalized to that of an ideal gas as a function of $T$ using the BN mass~(\ref{eq:PmD}) as well as $m_q=0$. For the strong coupling constant $\alpha_s$, we use three-loop running~\cite{Amsler:2008zzb} with $\Lambda_{\overline{\rm MS}}=344\,$MeV which for $N_f=3$ gives $\alpha_s({\rm 5\,GeV}) = 0.2034$~\cite{McNeile:2010ji}. The band is again obtained by varying the renormalization scale by a factor of 2 around the central value $\mu=2\pi T$. The $N_f = 2+1$ lattice data from the Wuppertal-Budapest collaboration use the stout action. Since their results show essentially no dependence on the lattice spacing (it is smaller than the statistical errors), they provide a continuum estimate by averaging the trace anomaly measured using their two smallest lattice spacings corresponding to $N_\tau = 8$ and $N_\tau = 10$~\cite{Borsanyi:2010cj}, which were essentially on top of the $N_\tau=6$ measurement \cite{Aoki:2005vt}.\footnote{It is noted that the Wuppertal-Budapest group has published a few data points  for the trace anomaly with $N_\tau =12$ and within statistical error bars these are consistent with the published continuum estimated results.} Using standard lattice techniques, the continuum-estimated pressure is computed from an integral of the trace anomaly. The $N_f = 2+1$ lattice data from the hotQCD collaboration are their $N_\tau = 8$ results using both the asqtad and p4 actions~\cite{Bazavov:2009zn}. The hotQCD results have not been continuum extrapolated and the error bars correspond to only statistical errors and do not factor in the systematic error associated with the calculation which, for the pressure, is estimated by the hotQCD collaboration to be between 5 - 10\%. As can be seen from the right panel of Fig.~\ref{fig:pressure}, the successive HTLpt approximations represent an improvement over the successive approximations coming from a weak-coupling expansion; however, as in the pure-glue case in the left panel of Fig.~\ref{fig:pressure}, the NNLO result represents a significant correction to the LO and NLO results. That being said, the NNLO HTLpt result agrees quite well with the available lattice data down to temperatures on the order of $2\,T_c \sim 340\,$MeV for QCD with $N_f=3$.\footnote{The Wuppertal-Budapest and hotQCD data were obtained using a physical strange quark mass; however, HTLpt calculations use massless quarks. The difference between massive and massless quarks is expected to be significant only for $T\lesssim32$\,MeV corresponding to the temperature where the lowest fermionic Matsubara mode equals the strange quark mass.} Below these temperatures the successive approximations give large corrections with the correction from NLO to NNLO reaching 100\% near $T_c$.

\subsubsection{$T^4$ scaled trace anomaly}

\begin{figure}[t]
\begin{center}
\includegraphics[width=0.45\textwidth]{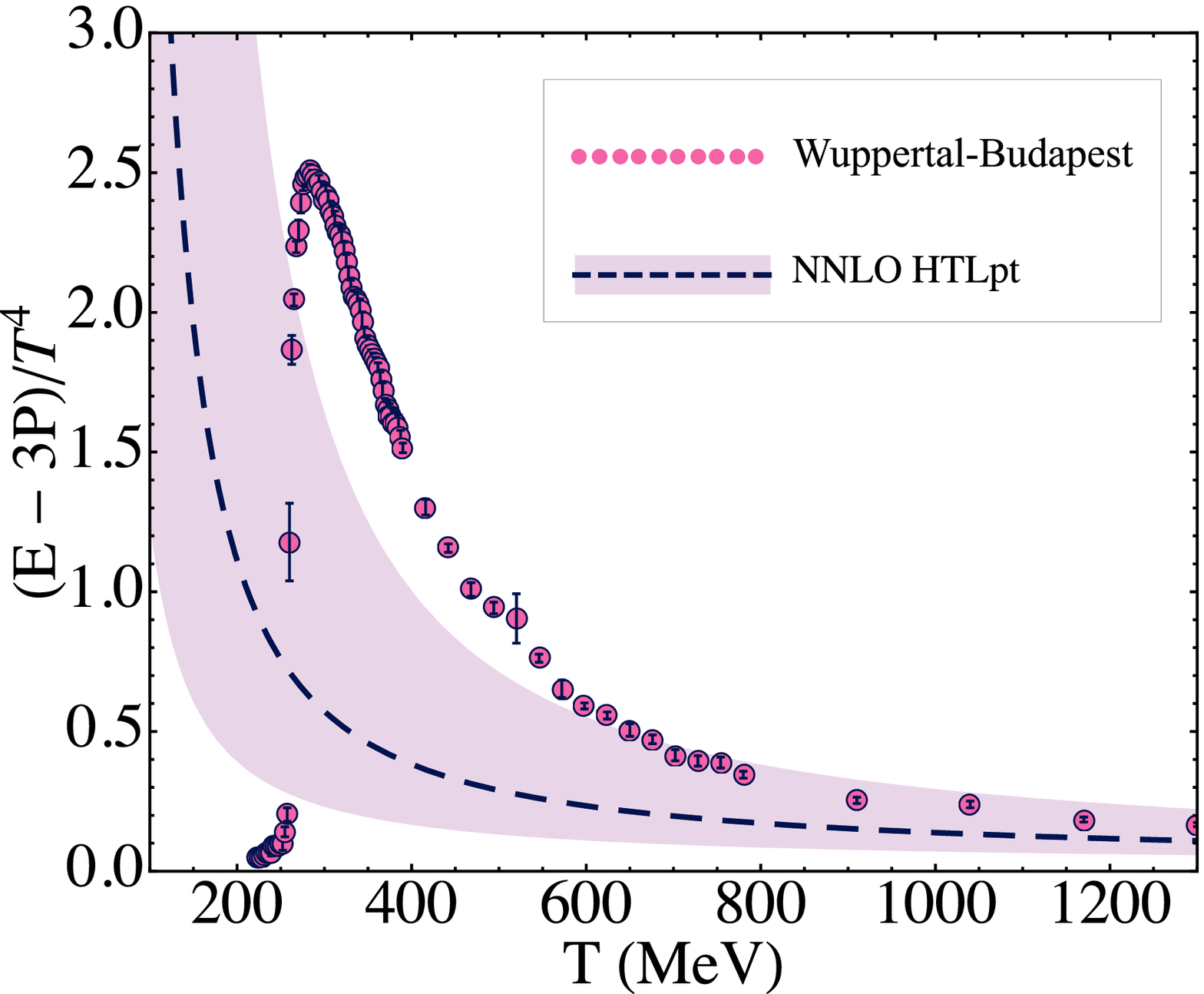}$\;\;\;\;$\includegraphics[width=0.46\textwidth]{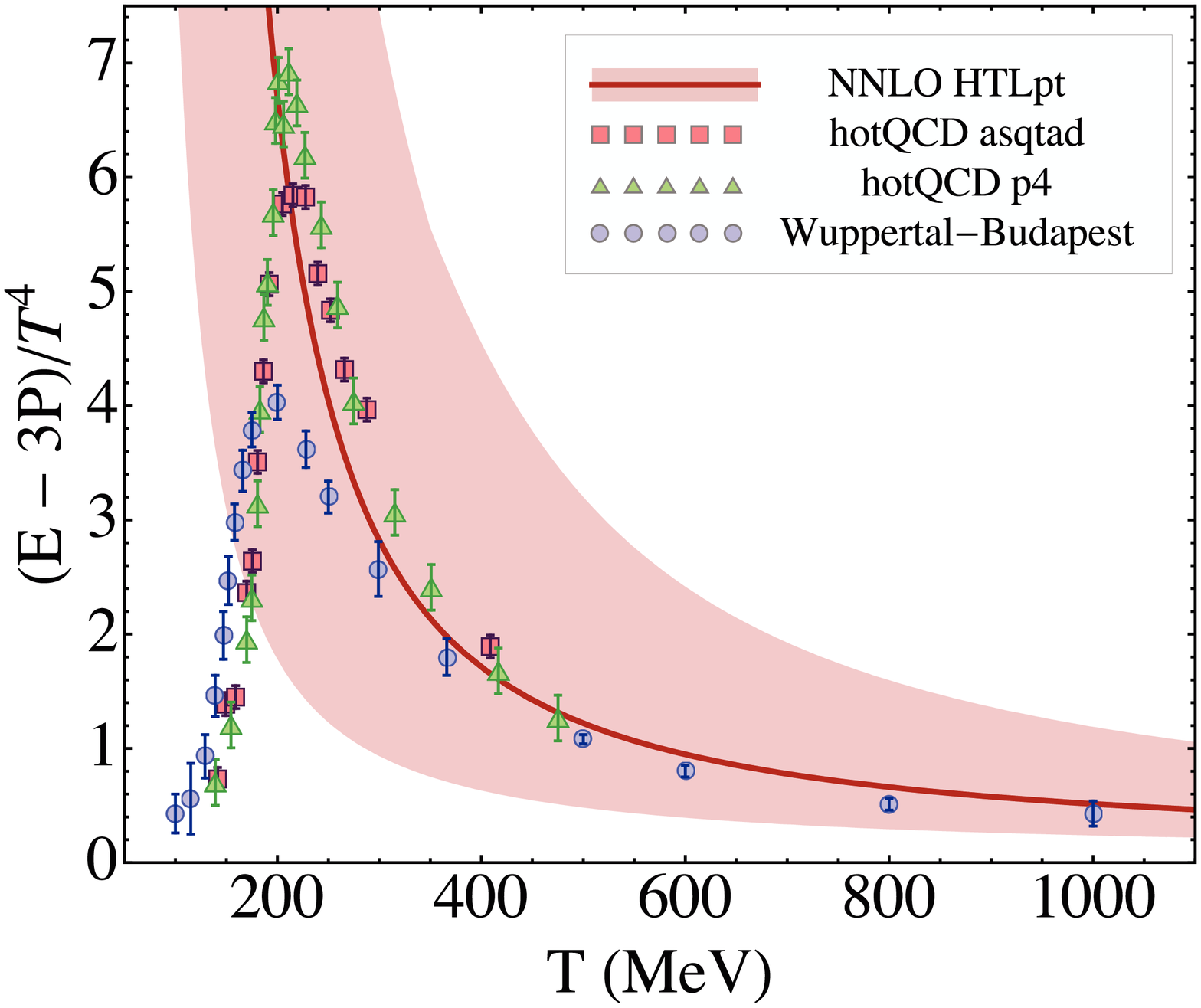}
\end{center}
\caption{\label{fig:anomaly} 
Comparison of NNLO predictions for the $T^4$ scaled trace anomaly for pure-glue QCD with lattice data from the Wuppertal-Budapest collaboration~\cite{Borsanyi:2011zm} (left panel) and $N_f=3$ QCD with $N_f = 2+1$ lattice data from the hotQCD~\cite{Bazavov:2009zn} and Wuppertal-Budapest~\cite{Borsanyi:2010cj} collaborations (right panel). Shaded band shows the result of varying the renormalization scale $\mu$ by a factor of 2 around $\mu = 2 \pi T$ for the NNLO result.}
\end{figure}

In the left panel of Fig.~\ref{fig:anomaly} we show the NNLO HTLpt prediction for the trace anomaly of pure-glue QCD normalized to $T^4$ as a function of $T$. The points are lattice data for pure-glue with $N_c=3$ from the Wuppertal-Budapest collaboration~\cite{Borsanyi:2011zm}. For temperatures below approximately $2\,T_c \sim 500\,$MeV, there is a large discrepancy between the HTLpt prediction and lattice data. The discrepancy decreases as increasing temperature, and for temperatures above approximately $4\,T_c \sim 1000\,$MeV, the NNLO HTLpt result is in good agreement with the lattice result.

In the right panel of Fig.~\ref{fig:anomaly} we show the NNLO HTLpt prediction for the trace anomaly of QCD with $N_f=3$ normalized to $T^4$ as a function of $T$. The data from both the Wuppertal-Budapest collaboration and the hotQCD collaboration are taken from the same data sets displayed in the right panel of Fig.~\ref{fig:pressure} and described previously. In the case of the hotQCD results we note that the results for the trace anomaly using the p4 action show large lattice size affects at all temperatures shown and the asqtad results for the trace anomaly show large lattice size effects for $T \gsim 200\,$MeV. We see very good agreement between the HTLpt prediction and the available lattice data down to temperatures on the order of $T \sim 2\,T_c$.

\subsubsection{$T^2$ scaled trace anomaly}

\begin{figure}[t]
\begin{center}
\includegraphics[width=0.47\textwidth]{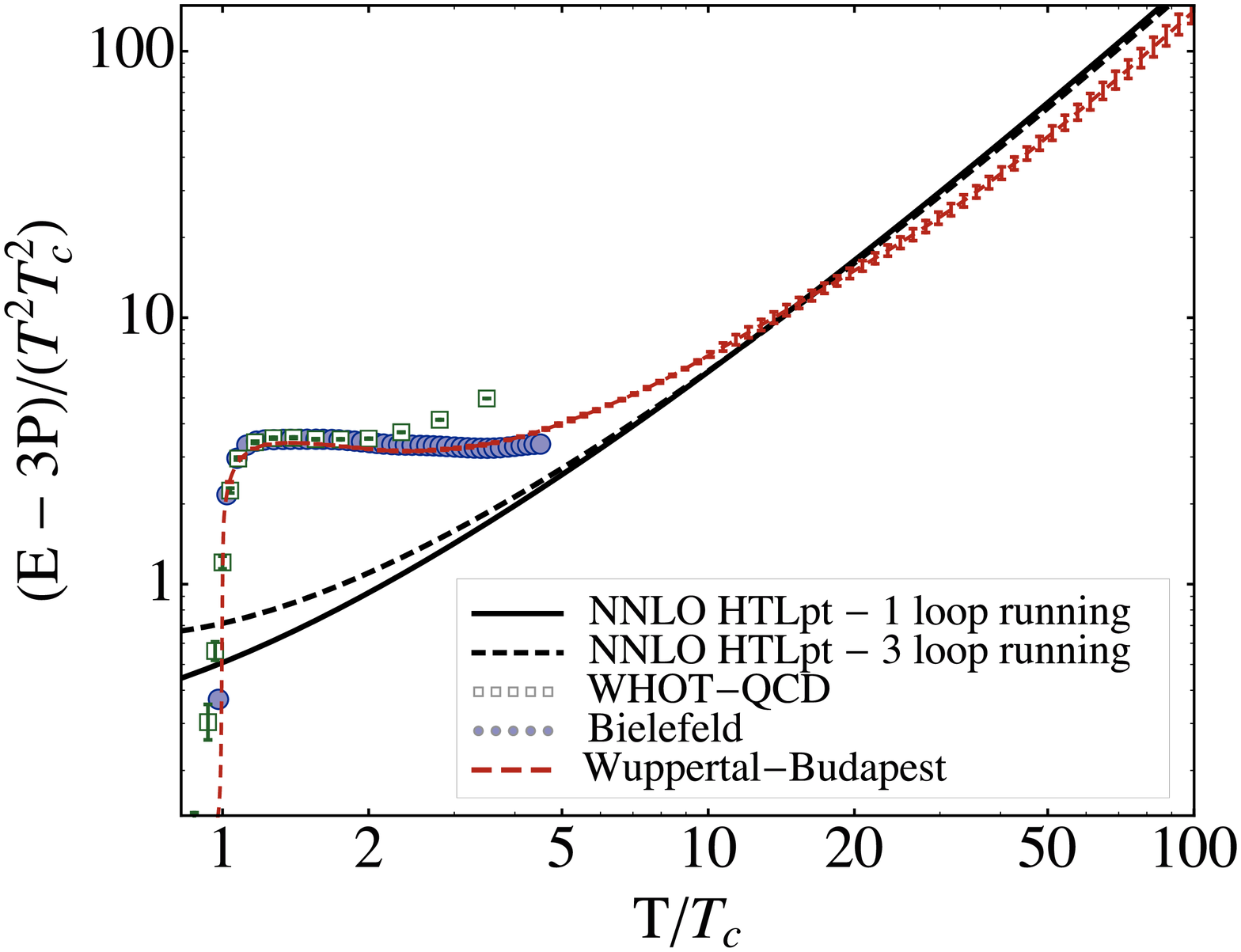}$\;\;\;\;$\includegraphics[width=0.47\textwidth]{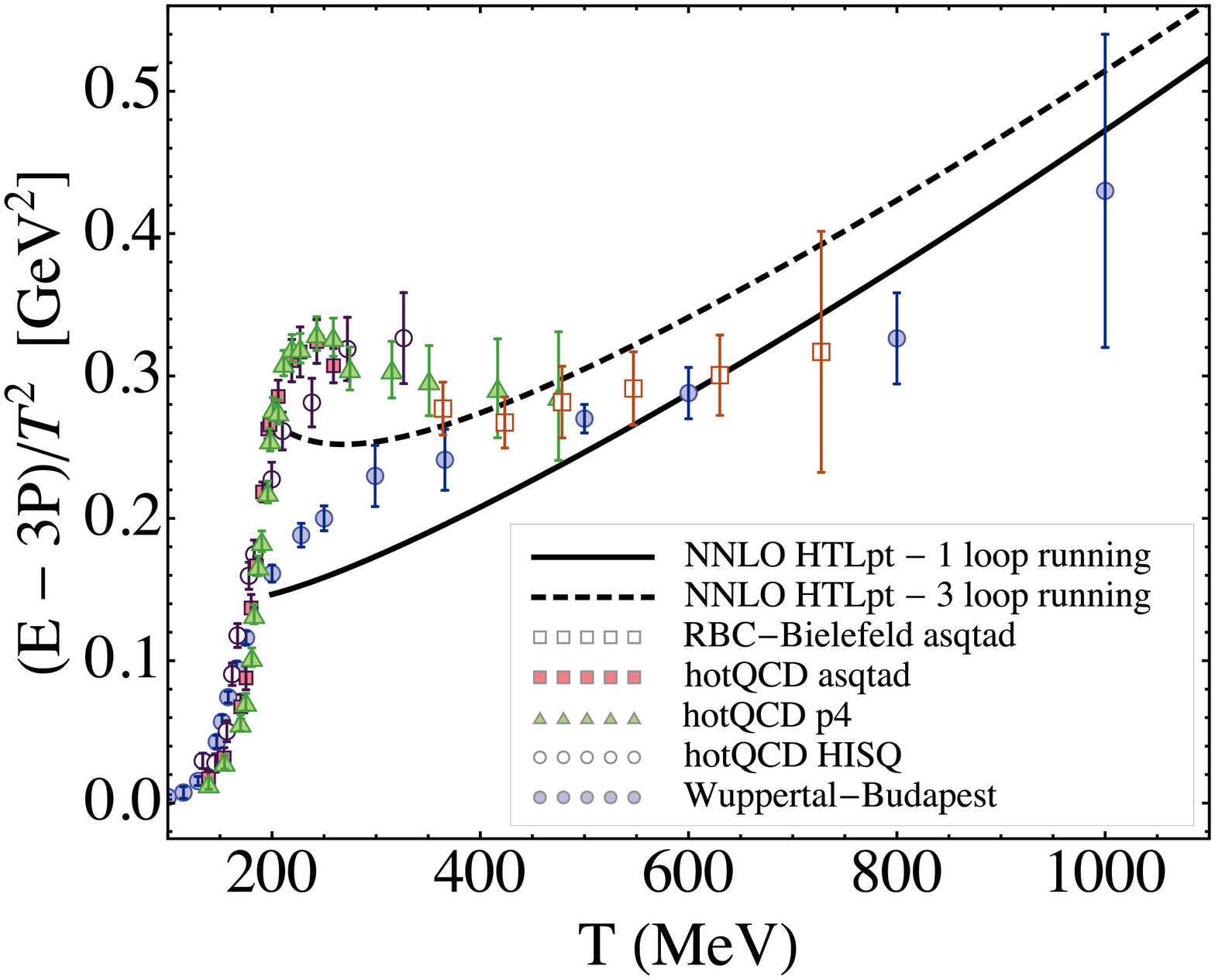}
\end{center}
\caption{\label{fig:anomaly-T2} 
Comparison of NNLO predictions for the $T^2$ scaled trace anomaly for pure-glue QCD with lattice data from the Wuppertal-Budapest~\cite{Borsanyi:2010cj}, Bielefeld~\cite{Boyd:1996bx}, and WHOT-QCD~\cite{Umeda:2008bd} collaborations (left panel) and $N_f = 3$ QCD with $N_f = 2+1$ lattice data from the Wuppertal-Budapest~\cite{Borsanyi:2010cj}, hotQCD~\cite{Bazavov:2009zn,Bazavov:2010pg}, and RBC-Bielefeld~\cite{Cheng:2007jq,Petreczky:2009at} collaborations (right panel). The HTLpt results are taken at $\mu=2\pi T$.}
\end{figure}

In the left panel of Fig.~\ref{fig:anomaly-T2}, we show comparison of the trace anomaly scaled by $T^2T_c^2$ for pure-glue QCD between NNLO HTLpt results and lattice data from the Wuppertal-Budapest~\cite{Borsanyi:2011zm}, Bielefeld~\cite{Boyd:1996bx}, and WHOT-QCD~\cite{Umeda:2008bd} collaborations. The solid black line is the NNLO HTLpt result obtained using a one-loop running coupling and the dashed black line is the HTLpt result obtained using a three-loop running  coupling. In the case of the three-loop running the lattice determination of $T_c/\Lambda_{\overline{\rm MS}} = 1.26$ is used to fix the scale \cite{Borsanyi:2011zm}. For comparison between the one- and three-loop results we require that both give the same value for the strong coupling when the renormalization scale $\mu = 5$ GeV. Numerically, one finds $\alpha_s(5\;{\rm GeV}) = 0.140553$. The difference between one- and three-loop running will be used to gauge the theoretical uncertainty of the NNLO HTLpt results. For both the one- and three-loop running $\mu = 2 \pi T$ is taken. As can be seen from the plot, at low temperatures NNLO HTLpt underpredicts the trace anomaly for pure-glue QCD and only starts to agree at temperatures on the order of $8\,T_c$. At high temperatures one sees excellent agreement between the NNLO HTLpt results and the lattice results. We note in this context that if one allows for a fit to the unknown perturbative $g^6$ contribution to the pressure, then the EQCD approach can equally-well describe the pure Yang-Mills lattice data down to temperatures on the order of $8\,T_c$ \cite{Hietanen:2008tv,Laine:2006cp}. The most remarkable feature of the lattice data for the $T^2$ scaled trace anomaly is that it is essentially constant in the temperature range $T \sim 1.1 - 4\,T_c$.  It has been suggested that this behavior is due to the influence of power corrections of order $T^2$ which are nonperturbative in nature and might be related to confinement~\cite{Kondo:2001nq,Meisinger:2001cq,Meisinger:2003id,Pisarski:2006hz,Pisarski:2006yk,Narison:2009ag}. At temperatures above approximately $4\,T_c$ the latest Wuppertal-Budapest results show an upward trend in accordance with perturbative predictions~\cite{Borsanyi:2011zm}. The WHOT-QCD results also exhibit an upward trend, however, it starts at much lower temperatures. This discrepancy could be due to their fixed scale approach not having sufficiently large $N_\tau$ at high temperatures as noted in their paper \cite{Umeda:2008bd}.

In the right panel of Fig.~\ref{fig:anomaly-T2}, we show the NNLO HTLpt trace anomaly scaled by $T^2$ for QCD with $N_f=3$ and compare to the $N_f = 2+1$ lattice results available from the Wuppertal-Budapest \cite{Borsanyi:2010cj}, hotQCD~\cite{Bazavov:2009zn,Bazavov:2010pg}, and RBC-Bielefeld~\cite{Cheng:2007jq,Petreczky:2009at} collaborations. As before, we show HTLpt results using both one- and three-loop running couplings with the requirement that both couplings give $\alpha_s({\rm 5\;GeV}) = 0.2034$ in accordance with recent high precision lattice measurements of the running coupling constant~\cite{McNeile:2010ji}. The lattice data from the Wuppertal-Budapest collaboration are taken from the same data sets displayed in Figs.~\ref{fig:pressure} and~\ref{fig:anomaly} and described previously. The lattice data from the hotQCD collaboration are their $N_\tau = 8$ results using the asqtad, p4, and HISQ actions which have not been continuum extrapolated \cite{Bazavov:2009zn,Bazavov:2010pg}. The lattice data from the RBC-Bielefeld collaboration is $N_\tau = 6$ and have also not been continuum extrapolated \cite{Cheng:2007jq,Petreczky:2009at}. As can be seen from the plot for $T > 400\,$MeV, one finds reasonable agreement between the NNLO HTLpt predictions and available lattice results. Below this temperature the Wuppertal-Budapest and hotQCD results do not seem to agree. Therefore, it is difficult to draw conclusions about the efficacy of the HTLpt approach; however, naively one expects that for temperatures less than twice the critical temperature $T_c$($\sim$170\,MeV for QCD with $N_f=2+1$) that nonperturbative corrections should become increasingly important.

\subsubsection{Discussions}

The HTLpt results indicate that the lattice data at temperatures $T \sim 2\,T_c$ are consistent with the quasiparticle picture. This is a nontrivial result since, in this temperature regime, the QCD coupling constant is neither infinitesimally weak nor infinitely strong with $g \sim 2$, or equivalently $\alpha_s \sim 0.4$. Therefore, we have a crucial test of the quasiparticle picture in the intermediate coupling regime.

The failure of HTLpt to match lattice data at lower temperatures is to be expected since one is expanding around the trivial vacuum $A_{\mu}=0$ and therefore neglects the approximate center symmetry $Z(N_c)$, which becomes essential as one approaches the deconfinement transition~\cite{KorthalsAltes:1999xb,KorthalsAltes:2000gs,Pisarski:2000eq,Vuorinen:2006nz,deForcrand:2008aw,Hidaka:2009hs,Dumitru:2010mj}. In addition, it is also in line with expectations since below $T\sim2-3\,T_c$ a simple ``electric'' quasiparticle approximation breaks down due to nonperturbative chromomagnetic effects~~\cite{Linde:1980ts,Gross:1980br}. Besides, there have been also hints that gauge-fixing ambiguities~\cite{Gribov:1977wm,Zwanziger:1988jt,Zwanziger:1989mf,Zahed:1999tg,Zwanziger:2004np,Zwanziger:2006sc,Lichtenegger:2008mh}, topological objects such as quantum instantons~\cite{Andersen:2006sf} and magnetic monopoles~\cite{Liao:2006ry,Liao:2008jg} might play important roles on the thermodynamics at intermediate temperature.

We find that when including quarks the agreement with lattice data is greatly improved as compared to the NNLO results of pure-glue QCD. Fermions are perturbative in the sense that they decouple in the dimensional-reduction step of effective field theory, so we expect that including contributions from quarks gives at least as good agreement with the lattice calculations as the pure-glue case. However, the exact reason for the better agreement between the HTLpt predictions and lattice calculations when including quarks is not clear to us.

Just as for NNLO QED and massless scalar $\phi^4$ theory, we encountered again the complex variational Debye mass when solving the gap equations. Whether the complexity of the variational Debye mass parameter is due to the additional expansion in $m_D/T$ and $m_q/T$ is impossible to decide at this stage. The correction to the pressure going from NLO to NNLO is also rather large. It is unfortunate that the nonperturbative magnetic scale prevents going to N$^3$LO without supplementing the calculation with input from three-dimensional lattice calculations, as it would be interesting to see whether the complexity of the Debye mass parameter and the slow convergence persists.

\section{Conclusions and outlook}
\label{concl}

In this paper, we have briefly reviewed the progress of hard-thermal-loop perturbation theory that has been made over the past 12 years concentrating on systematic thermodynamic calculations. We began by using the quantum mechanical anharmonic oscillator to show the breakdown of weak-coupling expansion which demonstrated, in sharp contrast to our intuition, that small coupling expansion may not be the same as perturbative expansion, or small coupling $\neq$ perturbative. Then we showed that as a resummation scheme, VPT was able to convert divergent weak-coupling expansions into fast converging approximations. With the inspiration from VPT, we introduced HTLpt as a reorganization for thermal gauge theory. The thermodynamic application of HTLpt leads to laudable results for both Abelian and non-Abelian theories.

The success of HTLpt is not totally unexpected since it is essentially just a reorganization of perturbation theory which shifts the expansion from around an ideal gas of massless particles to that of massive quasiparticles which are the more appropriate degrees of freedom at high temperature. The fact that the mass parameters are not arbitrary but functions of $g$ and $T$ determined variationally or perturbatively also indicates that HTLpt does not modify the original gauge theory but just reorganizes its perturbation series. Gauge invariance which is guaranteed by construction in HTLpt is useful both as a consistency check in calculations and as a way to simplify calculations. Although the renormalizability of HTLpt is not yet proven, the fact that it is renormalizable at NNLO using only known counterterms shows promising light along the way.

So far, thermodynamics for quantum fields has been studied intensively in the community, both perturbatively through higher orders or numerically on the lattice, however real-time dynamics is still in its very early stage of development. Transport coefficients are of great interest since they are theoretically clean and well defined non-equilibrium dynamical quantities. Along the line of perturbative approach to transport coefficients, the only known ones to NLO are shear viscosity in scalar $\phi^4$ theory~\cite{Moore:2007ib}, heavy quark diffusion in QCD and ${\cal N}=4$ supersymmetric Yang-Mills theory~\cite{CaronHuot:2007gq,CaronHuot:2008uh}, and transverse diffusion rate $\hat{q}$ in QCD~\cite{CaronHuot:2008ni}, and all of them exhibit poor convergence as bad as the case of thermodynamic quantities, such as the pressure. Since dynamical quantities are still not well described by lattice gauge theory, new resummation techniques are urgently needed in order to achieve a better understanding of transport coefficients.

Although the papers written to date have focussed on using HTLpt to compute thermodynamic observables, the goal of this project is to create a framework which can be applied to both equilibrium and non-equilibrium systems. HTLpt is formulated in Minkowski space, so its application to real-time dynamics is straightforward. With the confidence from thermodynamic calculations, HTLpt is ready to enter the domain of real-time dynamics at temperatures that are relevant for LHC and this might be of great help in deepening our knowledge in the properties of the quark-gluon plasma. Last but not least, it would be also interesting to explore the applicability of the HTLpt/SPT method to other finite temperature or density systems, such as compact stars~\cite{Andersen:2002jz} and ultracold atoms~\cite{Braaten:2002fc,Braaten:2002zz}.

\section*{Acknowledgments}

The author would like to thank Jens O. Andersen, Lars E. Leganger, and Michael Strickland for a fruitful collaboration. The author acknowledges support from the Alexander von Humboldt Foundation.
\appendix

\section{HTLpt diagrams through NNLO}
\label{diagr}

\begin{figure}[t]\centering
\includegraphics[trim=102px 240px 102px 66px,clip]{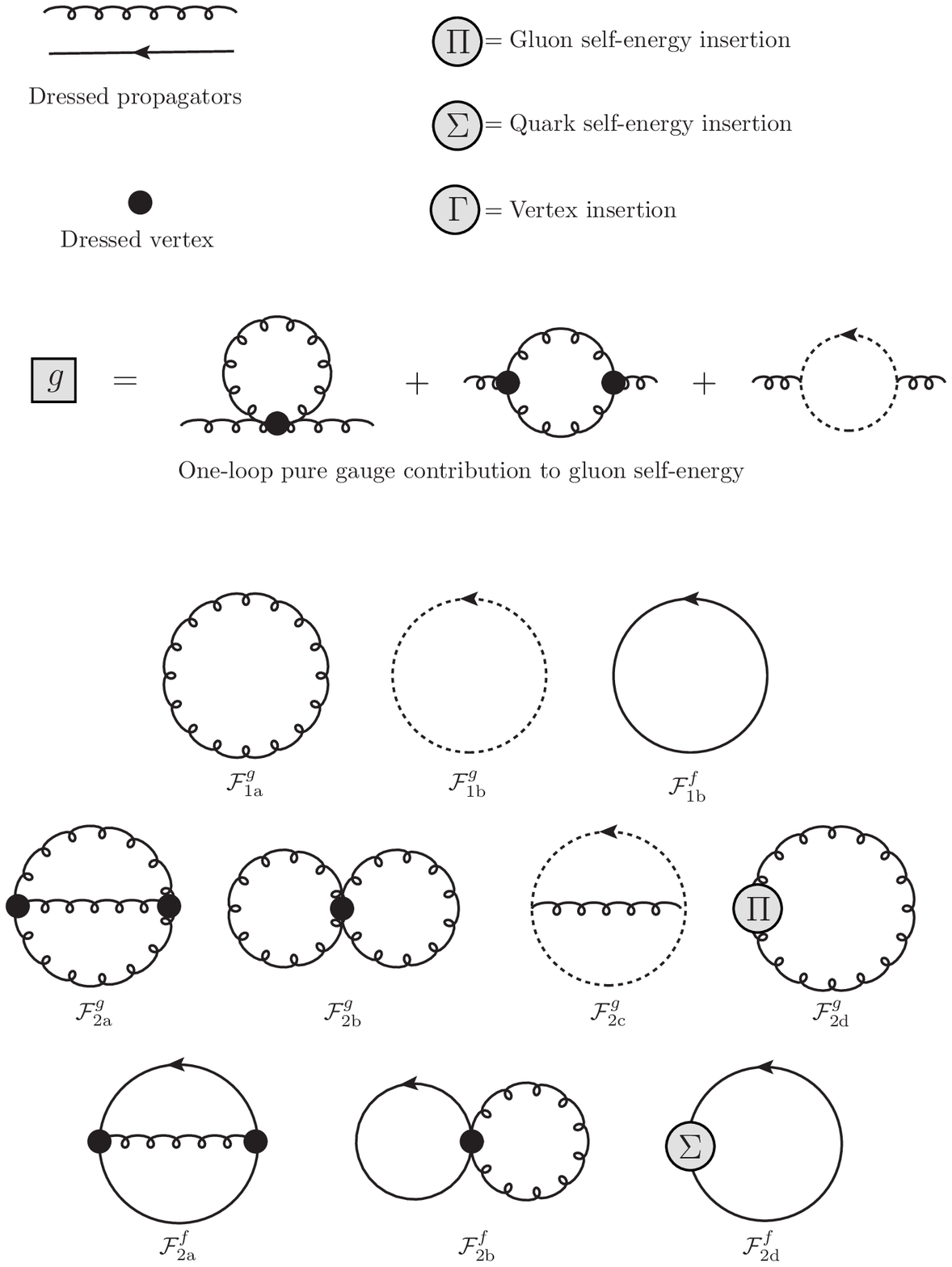}
\caption{QCD diagrams contributing through NLO in HTLpt. The spiral lines are gluon propagators, the dotted lines are ghost propagators, and the solid lines are quark propagators. A circle with a $\Pi$ indicates a one-loop gluon self-energy insertion, a $\Sigma$ indicates a one-loop quark self-energy insertion, and a $\Gamma$ indicates a one-loop vertex insertion. A square with a $g$ is shorthand for the pure-glue diagrams contributing to the one-loop gluon self-energy. All gluon and quark propagators and vertices shown are HTL-resummed propagators and vertices. The logic behind the diagram notation is as follows: diagrams consisting only of gauge propagators have $g$ superscripts. Diagrams containing fermion propagators have $f$ superscripts. The subscript indices are identical to those used in \cite{Andersen:2010ct} (pure-glue QCD) and \cite{Andersen:2009tw} (QED). We do not display the symmetry factors in the diagrams.}
\label{fig:LO&NLO}
\end{figure}

\begin{figure}[t]\centering
\includegraphics[trim=104px 169px 95px 23px,clip]{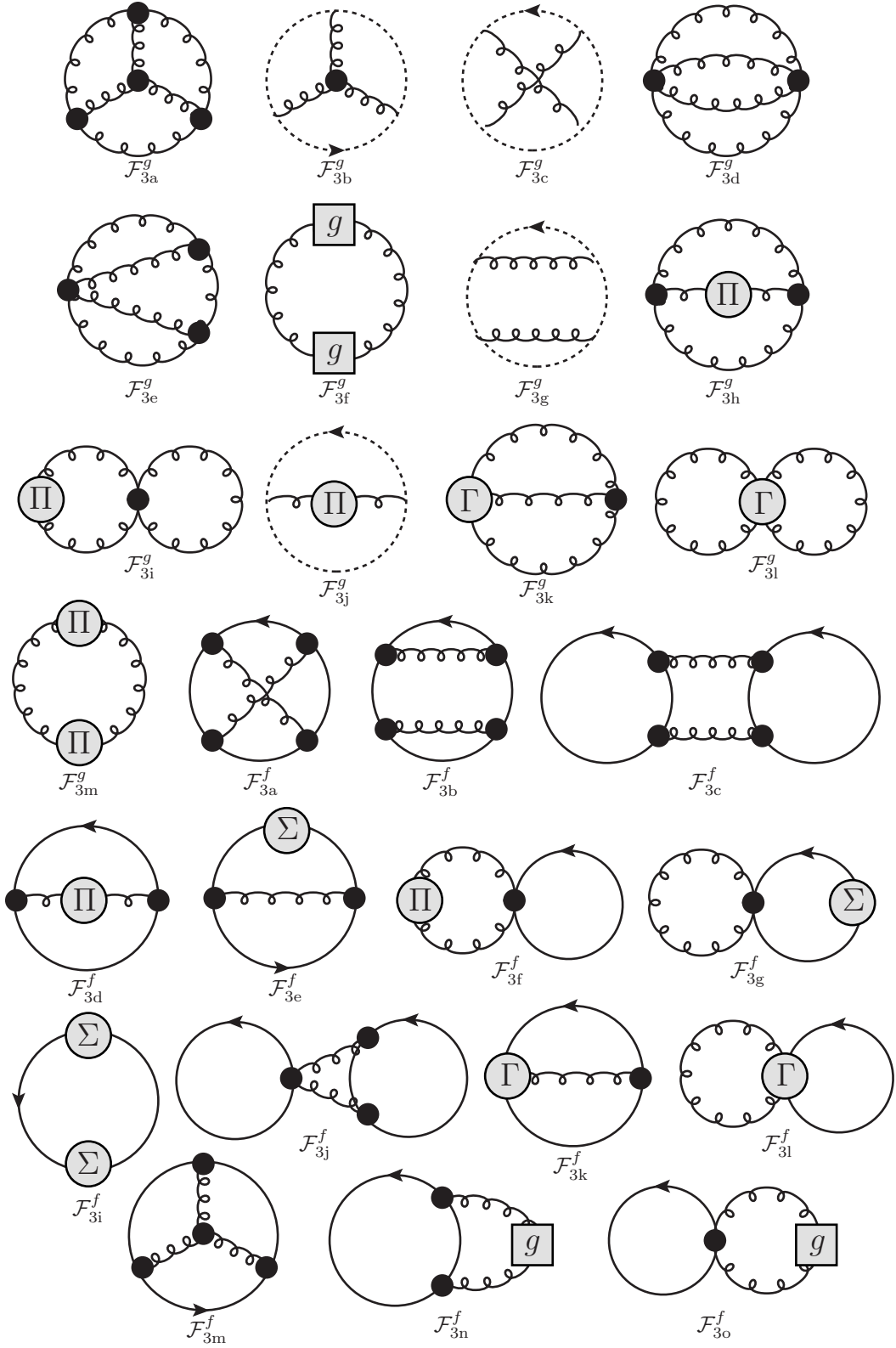}
\caption{QCD diagrams contributing at NNLO in HTLpt.}
\label{fig:NNLO}
\end{figure}

\bibliographystyle{apsrev4-1}
\bibliography{Su.bib}

\end{document}